\documentclass[aps,pra,twocolumn,amsmath,amssymb,nofootinbib,superscriptaddress]{revtex4}

\newcommand{\bra}[1]{\langle#1|}
\newcommand{\ket}[1]{|#1\rangle}

\usepackage[pdftex]{graphicx}
\usepackage{mathrsfs}
\usepackage[colorlinks]{hyperref}

\begin{document}

\bibliographystyle{apsrev}

%
%

\title{Boson-sampling with photons of arbitrary spectral structure}

%
%

\author{Peter P. Rohde}
\email[]{dr.rohde@gmail.com}
\homepage{http://www.peterrohde.org}
\affiliation{Centre for Engineered Quantum Systems, Department of Physics and Astronomy, Macquarie University, Sydney NSW 2113, Australia}

\date{\today}

\frenchspacing

%
%

\begin{abstract}
Boson-sampling has attracted much interest as a simplified approach to implementing a subset of optical quantum computing. Boson-sampling requires indistinguishable photons, but far fewer of them than universal optical quantum computing architectures. In reality, photons are never indistinguishable, and exhibit a rich spectral structure. Here we consider the operation of boson-sampling with photons of arbitrary spectral structure and relate the sampling statistics of the device to matrix permanents. This sheds light on the computational complexity of different regimes of the photons' spectral characteristics, and provides very general results for the operation of linear optics interferometers in the presence of partially distinguishable photons. Our results apply to both the cases of spectrally resolving and non-spectrally resolving detectors.
\end{abstract}

\maketitle

%
%

\section{Introduction}

Linear optics interferometry has widespread uses in optical quantum computing \cite{bib:KLM01, bib:KokLovett11, bib:NielsenChuang00}, quantum metrology, and quantum cryptography. In particular, passive linear optics interferometry is the basis of the recent field of boson-sampling \cite{bib:AaronsonArkhipov10}, where a series of single photons are evolved via linear optics and subsequently sampled using coincidence photodetection, a problem which has been shown to be classically intractable.

Such interferometry typically requires indistinguishable photonic states, such that generalized Hong-Ou-Mandel (HOM) \cite{bib:HOM87, bib:RohdeRalph05} interference takes place. In boson-sampling, it was found that the amplitudes in the system are related to matrix permanents. On the other hand, de Guise \emph{et al.} \cite{bib:SandersImm} and Tillmann \emph{et al.} \cite{bib:TillmannImm14} showed that linear optics networks with time delays relates to matrix immanents.

Here we generalize the boson-sampling model to the situation where the photons have arbitrary spectral structure \cite{bib:RohdeArb} and show how the sampling probabilities relate to functions of matrix permanents. This sheds light on the question `how does the spectral structure of photons relate to their computational complexity?'. Our results reconfirm the expectation that indistinguishable photonic systems should reduce to ideal boson-sampling, which is computationally complex, whereas distinguishable photons are computationally trivial. Our results provide general expressions for the behavior of such systems in the intermediate regimes with arbitrary, non-identical spectral structures. We consider both spectrally resolved and the more realistic non-spectrally resolved detectors. As an elementary demonstration of the techniques, we reproduce HOM interference in both detector regimes.

%
%

\section{Boson-sampling}

In the boson-sampling model, we begin by preparing $n$ single photons in $m$ modes,
\begin{eqnarray} \label{eq:ideal_input}
\ket{\psi_\mathrm{in}} &=& \hat{a}^\dag_1 \dots \hat{a}^\dag_n \ket{0_1,\dots,0_m} \nonumber \\
&=& \prod_{i=1}^n \hat{a}_i^\dag \ket{\vec{0}},
\end{eqnarray}
where \mbox{$\hat{a}^\dag_i$} is the photonic creation operator on the $i$th mode. This state is evolved via passive linear optics (i.e beamsplitters and phase-shifters), which implements the unitary transformation,
\begin{equation}
\hat{U}\hat{a}_i^\dag\hat{U}^\dag \to \sum_{j=1}^m U_{i,j} \hat{a}_j^\dag.
\end{equation}
The output state of the system is of the form,
\begin{equation}
\ket{\psi_\mathrm{out}} = \sum_S \gamma_S \ket{S_1,\dots,S_m},
\end{equation}
where $S$ are the photon number configurations and $S_i$ is the number of photons in the $i$th mode associated with configuration $S$. Scheel \cite{bib:Scheel04} found that the amplitudes $\gamma_S$ are related to matrix permanents as,
\begin{equation}
\gamma_S = \frac{\mathrm{Per}(U_{S,T})}{\sqrt{S_1!\dots S_m! T_1!\dots T_m!}},
\end{equation}
where $U_{S,T}$ is an \mbox{$n\times n$} sub-matrix of $U$ as a function of the input ($T$) and output ($S$) configurations. $U_{S,T}$ is obtained as follows. Let \mbox{$S=\{S_1,\dots,S_m\}$} and \mbox{$T=\{T_1,\dots,T_m\}$}, where $S_i$ and $T_i$ are the number of photons in the $i$th mode of the respective configuration. Then we define the \mbox{$n\times m$} submatrix $U_T$ by taking $T_j$ copies of the $j$th column of $U$ for each $j$. Then $U_{S,T}$ is obtained by taking $S_i$ copies of the $i$th row of $U_T$ for each $i$ to obtain the \mbox{$n\times n$} submatrix $U_{S,T}$.

The computational complexity of boson-sampling relates to the fact that calculating the permanents of complex-valued matrices is \textbf{\#P}-complete, a complexity class believed to be classically hard to simulate.

The number of modes in a boson-sampling device scales as \mbox{$m=O(n^2)$}. Thus, for large systems we are likely to never have more that one one photon at a given output mode. This is the binary regime, whereby every mode has 0 or 1 photons, and we will make this assumption throughout to simplify notation.

For an elementary introduction to boson-sampling, see Gard \emph{et al.} \cite{bib:GardBSintro}. And for a complete description, including the full complexity proof, refer to Aaronson \& Arkhipov \cite{bib:AaronsonArkhipov10}.

%
%

\section{Spectral structure of photons}

Photons exhibit rich spectral structure, and a representation in terms of \mbox{$\hat{a}^\dag$} does not characterize this. Instead we will adopt the mode operator formalism \cite{bib:RohdeMauererSilberhorn07}, whereby photons are represented by mode operators of the form,
\begin{equation}
\hat{A}_{\psi,j}^\dag = \int \psi(\omega) \hat{a}_j^\dag(\omega)\, \mathrm{d}\omega,
\end{equation}
where \mbox{$\hat{A}_{\psi,j}^\dag$} creates a photon with spectral distribution function \mbox{$\psi(\omega)$} in the $j$th mode, \mbox{$\hat{a}^\dag(\omega)$} is the photonic creation operator at frequency $\omega$, and the integral is over all frequencies. To satisfy normalization, we require,
\begin{equation}
\int |\psi(\omega)|^2 \,\mathrm{d}\omega = 1.
\end{equation}
Next, we may choose a discrete orthonormal basis in which to express \mbox{$\psi(\omega)$},
\begin{equation}
\psi(\omega) = \sum_i \lambda_i \xi_i(\omega),
\end{equation}
where $\lambda$ are the coefficients in the decomposition and $\xi$ are the basis functions. The $\lambda$ coefficients may be calculated using,
\begin{equation}
\lambda_i = \int \xi_i(\omega)^* \psi(\omega)\, \mathrm{d}\omega.
\end{equation}
In order for $\xi_i$ to be a valid basis, we require,
\begin{equation} \label{eq:orthonormality}
\int \xi_i(\omega)^*\xi_j(\omega)\,\mathrm{d}\omega = \bra{0} \hat{A}_{\xi_i} \hat{A}_{\xi_j}^\dag \ket{0} = \delta_{i,j}.
\end{equation}
It follows that the mode operators can be expressed as a decomposition into an orthonormal basis of mode operators,
\begin{eqnarray}
\hat{A}_{\psi,j}^\dag &=& \sum_i \lambda_i \int \xi_i(\omega) \hat{a}_j^\dag(\omega)\, \mathrm{d}\omega \nonumber \\
&=& \sum_i \lambda_i \hat{A}_{\xi_i,j}^\dag.
\end{eqnarray}

%
%

\section{Boson-sampling with arbitrary spectral states}

\subsection{Spectrally pure photons}

Let the input state to the interferometer be a tensor product of $n$ photons, each characterized by distinct spectral distribution functions,
\begin{eqnarray}
\ket{\psi_\mathrm{in}} &=& \prod_{j=1}^n \hat{A}_{\psi_j,j}^\dag \ket{\vec{0}} \nonumber \\
&=& \prod_{j=1}^n \sum_i \lambda_{i,j} \hat{A}_{\xi_i,j}^\dag \ket{\vec{0}},
\end{eqnarray}
This may be re-expressed as \cite{bib:RohdeLowFid12},
\begin{equation} \label{eq:simp_spectral_input}
\ket{\psi_\mathrm{in}} = \sum_{v\in V} \left(\prod_{j=1}^n \lambda_{v_j,j} \cdot \prod_{j=1}^n \hat{A}_{\xi_{v_j},j}^\dag \right) \ket{\vec{0}},
\end{equation}
where $V$ is the set of all vectors of length $n$ with integer indices spanning the support of the discrete basis, which we let be $N$. Each $v$ can be interpreted as a configuration of spectral modes at the input. For example, \mbox{$v=\{1,1,2,3\}$} means that the first and second modes are in spectral basis function $\xi_1$, the third mode is in $\xi_2$, and the fourth mode in $\xi_3$. Note that the number of spectral configurations, $|V|$, is exponential in $n$ (unless all photons are indistinguishable, in which case trivially \mbox{$|V|=1$}, and Eq. \ref{eq:simp_spectral_input} reduces to the ideal Eq. \ref{eq:ideal_input}).

Then,
\begin{equation}
\ket{\psi_\mathrm{in}} = \sum_{v\in V} \chi_{\vec\psi}(v) \prod_{i=1}^N \prod_{j\in T(v,i)} \hat{A}_{\xi_i,j}^\dag \ket{\vec{0}},
\end{equation}
where,
\begin{equation}
\chi_{\vec\psi}(v) = \prod_{j=1}^n \lambda_{v_j,j},
\end{equation}
and \mbox{$T(v,i)$} is the set of spatial modes for spectral configuration $v$ where the elements of $v$ are $i$. Here \mbox{$\vec\psi=\{\psi_1,\dots,\psi_n\}$} denotes the set of spectral distribution functions of each of the input photons.

Next we apply the linear optics evolution, which transforms the input state to,
\begin{equation}
\ket{\psi_\mathrm{out}} = \sum_{v\in V} \Bigg(\chi_{\vec\psi}(v) \prod_{i=1}^N \underbrace{\prod_{j\in T(v,i)} \hat{U} \hat{A}_{\xi_i,j}^\dag \hat{U}^\dag}_{|T(v,i)|\,\mathrm{indist.\,photons}} \Bigg) \ket{\vec{0}}.
\end{equation}
The underbraced component of the equation represents a set of $k=|T(v,i)|$ indistinguishable photons in spectral mode $\xi_i$, and evolves using the standard permanent rule for indistinguishable photons,
\begin{equation} \label{eq:generalOut}
\ket{\psi_\mathrm{out}} = \sum_{v\in V} \chi_{\vec\psi}(v) \prod_{i=1}^N \sum_S \mathrm{Per}\left(U_{S,T(v,i)}\right) \prod_{j=1}^m \left(\hat{A}^\dag_{\xi_i,j}\right)^{S_j} \ket{\vec{0}},
\end{equation}
where $S$ is the set of all allowed output configurations of $k$ photons, as before. Here \mbox{$U_{S,T(v,i)}$} is a \mbox{$k\times k$} sub-matrix of $U$, obtained using the usual linear optics permanent rule.

%
%

\subsection{Spectrally-resolved detectors}

Let us first assume that our photodetectors are able to distinguish the different spectral basis functions. If we know \emph{a priori} what the spectral characteristics of the detector are, we can choose the spectral decomposition basis correspondingly. Let $S^{(i)}$ be the configuration associated with the $i$th spectral mode. If the detector can project uniquely onto spectral mode $\xi_i$ then the associated amplitude is,
\begin{equation} \label{eq:spectral_amplitude_resolved}
\gamma_{\vec\psi}(S^{(i)}) = \sum_{v\in V} \chi_{\vec\psi}(v) \mathrm{Per}\left(U_{S^{(i)},T(v,i)}\right).
\end{equation}
Clearly, if the photons are all indistinguishable (i.e all photons are described by the same spectral decomposition), then \mbox{$|V|=1$} (when decomposed into an appropriate basis) and Eq. \ref{eq:spectral_amplitude_resolved} reduces to normal permanent sampling. Otherwise \mbox{$|V|>1$}, and we are sampling from linear combinations of permanents.

Let \mbox{$\vec{S}=\{S^{(1)},\dots,S^{(N)}\}$} denote the configuration across all spectral modes, where again $S^{(i)}$ is the configuration associated with the $i$th spectral mode. Then the associated amplitude is,
\begin{equation} \label{eq:general_gamma}
\gamma_{\vec\psi}(\vec{S}) = \sum_{v\in V} \chi_{\vec\psi}(v) \prod_{i=1}^N \mathrm{Per}\left(U_{S^{(i)},T(v,i)}\right),
\end{equation}
and the respective measurement probability is \mbox{$P_{\vec\psi}(\vec{S})=|\gamma_{\vec\psi}(\vec{S})|^2$}.

%
%

\subsection{Non-spectrally-resolved detectors}

In reality, photodetectors are typically unable to resolve an orthonormal basis of spectral functions, with exception to time-resolved photodetection where the response of the detector is much shorter than the length of the wavepacket, or detectors with narrowband frequency filtering. Next we will consider the situation where non-resolving detectors are employed. That is, the detectors can tell us how many photons arrived, but nothing about their spectral structure.

Let $M$ be a measurement signature, which is a configuration outcome, irrespective of the spectral modes in which the photons were measured. They satisfy the constraint $\sum_i S^{(i)} = M$. Then the probability of that measurement signature is given by summing over all partitions of the measurement signature into signatures within individual spectral modes,
\begin{eqnarray} \label{eq:generalPM}
P(M) &=& \sum_{\vec{S}\,\,\mathrm{s.t.}\,\sum_i S^{(i)} = M} P(\vec{S}) \nonumber \\
&=& \underbrace{\sum_{\vec{S}}}_{\mathrm{classical}} \Bigg| \underbrace{\sum_{v\in V} \chi_{\vec\psi}(v) \prod_{i=1}^N \mathrm{Per}\left(U_{S^{(i)},T(v,i)}\right)}_{\mathrm{quantum}}\Bigg|^2,\nonumber \\
\end{eqnarray}
where we sum over all signatures $S$ associated with each spectral mode, such that the sum of the signatures is the measurement result $M$. Eq. \ref{eq:generalPM} is structurally similar to a permanent of permanents. The first permanent-like function (the sum over $\vec{S}$), labeled `classical', is of positive real-valued elements. This term sums the classically distinct elements -- the different spectral basis functions that the detector is unable to resolve. The second component, labeled `quantum', contains the permanents associated with the sampling of each individual spectral component associated with the respective $\vec{S}$. 

%
%

\subsection{Limiting cases}

The $\lambda$ matrix characterizes the spectral decomposition across all modes. Two limiting cases are of particular interest. When all of the photons are indistinguishable, they all reside in the same spectral mode, and,
\begin{equation}
\lambda =
\left[ \begin{array}{cccc}
1 & 0 & 0 & \ldots \\
1 & 0 & 0 & \ldots \\
1 & 0 & 0 & \ldots \\
\vdots & \vdots & \vdots & \ddots
\end{array} \right].
\end{equation}
In this instance our parameters are given by,
\begin{itemize}
\item \mbox{$\chi(\{1,1,1,\dots\})=1$}, otherwise \mbox{$\chi(v)=0$}.
\item \mbox{$T(\{1,1,1,\dots\},1)=\{1,2,3,\dots\}$}.
\item $S^{(1)}=M$, otherwise $S^{(j)}=\{\}$
\end{itemize}
Now Eq. \ref{eq:generalPM} reduces to,
\begin{equation}
P(M) = \left|\mathrm{Per}\left(U_{M,\{1,2,3,\dots\}}\right)\right|^2,
\end{equation}
which is the expected boson-sampling result, where the permanent is of an \mbox{$n\times n$} submatrix of $U$.

Alternately, when all photons are distinguishable, there is no overlap between their spectral coefficients and,
\begin{equation}
\lambda =
\left[ \begin{array}{cccc}
1 & 0 & 0 & \ldots \\
0 & 1 & 0 & \ldots \\
0 & 0 & 1 & \ldots \\
\vdots & \vdots & \vdots & \ddots
\end{array} \right].
\end{equation}
Now the parameters are given by,
\begin{itemize}
\item \mbox{$\chi(\{1,2,3,\dots\})=1$}, otherwise \mbox{$\chi(v)=0$}
\item \mbox{$T(\{1,2,3,\dots\},i)=\{i\}$}
\item Since all photons are distinguishable and are no longer permutation symmetric, the allowed signatures $\vec{S}$ are all the permutations of single photons reaching the respective outputs, giving rise to $n!$ terms.
\end{itemize}
Thus,
\begin{eqnarray}
P(M) &=& \sum_{x\in \sigma_n} \prod_{i=1}^N \left|\mathrm{Per}\left(U_{\{x_i\},\{i\}}\right)\right|^2 \\ \nonumber
&=& \sum_{x\in \sigma_n} \prod_{i=1}^N \left|U_{\{x_i\},\{i\}}\right|^2 \\ \nonumber
&=& \mathrm{Per}\left(\left|U_{M,\{1,2,3,\dots\}}\right|^2\right),
\end{eqnarray}
where $\sigma_n$ are the permutations of $n$ elements over the modes in the configuration, and the square is element-wise. Note that the permanent is of a positive, real-valued matrix (a classical probability distribution), which is computationally easy to approximate \cite{bib:SinclairPerm}. This is expected for distinguishable photons, since each photon's evolution can be evaluated independently. Intuitively, this equation tells us that the sampling probabilities are related to the combinatorics of classical probabilities, which is expected since none of the photons interfere.

The only difference between the two examples (completely indistinguishable and completely distinguishable) is the location of the absolute square. When dealing with indistinguishable photons the permanent is of quantum amplitudes, and the permanent is absolute squared to yield a classical probability, whereas for distinguishable photons we take the permanent of classical probabilities, since the photons do not interfere.

In the intermediate regime we will be evaluating a combinatoric expression over permanents of matrices varying in size from $1$ to $n$.

%
%

\subsection{Hong-Ou-Mandel interference}

The simplest example to consider is HOM interference, of two photons at a 50/50 beamsplitter, given by the Hadamard matrix,
\begin{equation}
U = \frac{1}{\sqrt{2}} \left[ \begin{array}{cc}
1 & 1 \\
1 & -1 \\
\end{array} \right].
\end{equation}
Let the incident photons be a superposition across two spectral modes. Then our spectral decomposition matrix is given by,
\begin{equation}
\lambda = \left[ \begin{array}{cc}
1 & 0 \\
\alpha & \sqrt{1-\alpha^2} \\
\end{array} \right],
\end{equation}
where \mbox{$0\leq \alpha\leq 1$} characterizes the distinguishability of the two photons (\mbox{$\alpha=1$} for indistinguishable photons, and \mbox{$\alpha=0$} for distinguishable photons). Now $V$ is the set of all length-2 vectors with indices from 1 to 2: \mbox{$v_1=\{1,1\}$}, \mbox{$v_2=\{1,2\}$}, \mbox{$v_3=\{2,1\}$}, \mbox{$v_4=\{2,2\}$}. This gives: \mbox{$\chi(v_1)=\alpha$}, \mbox{$\chi(v_2)=\sqrt{1-\alpha^2}$}, \mbox{$\chi(v_3)=\chi(v_4)=0$}. And, \mbox{$T(v_1,1)=\{1,2\}$}, \mbox{$T(v_1,2)=\{\}$}, \mbox{$T(v_2,1)=\{1\}$}, \mbox{$T(v_2,2)=\{2\}$}.

First let us consider the case where the detector is spectrally-resolving and can project onto the $\xi_1$ spectral mode. This might be implemented experimentally using narrowband filtering, if the two spectral basis states were frequency eigenstates. Let \mbox{$S^{(1)}=\{1,1\}$} be the anti-bunched signature in the first spectral mode, where both photons were found at different output ports. Then we have,
\begin{equation}
\gamma(S^{(1)}) = \alpha \cdot\mathrm{Per}\left(U_{\{1,1\},\{1,1\}}\right) = 0,
\end{equation}
and we never observe anti-bunching. This is the expected result for HOM interference using indistinguishable photons (or partially distinguishable photons with spectral filtering).

On the other hand, let us consider the case where the detectors are not spectrally-resolving, and are `blind' to the spectral structure of the photons. Then our measurement outcome for the anti-bunched case is \mbox{$M=\{1,1\}$}, and the allowed partitions of $M$ into individual spectrally-resolved signatures are: \mbox{$S^{(1)}=\{1,0\}$}, \mbox{$S^{(2)}=\{0,1\}$}; or, \mbox{$S^{(1)}=\{0,1\}$}, \mbox{$S^{(2)}=\{1,0\}$}; or, \mbox{$S^{(1)}=\{1,1\}$}, \mbox{$S^{(2)}=\{0,0\}$}. It is then easily seen that,
\begin{eqnarray}
P(M) &=& P(\{1,0\},\{0,1\}) \nonumber \\
&+& P(\{0,1\},\{1,0\}) \nonumber \\
&+& P(\{1,1\},\{0,0\}) \\ \nonumber
&=& \frac{1-\alpha^2}{4} + \frac{1-\alpha^2}{4} + 0 \nonumber \\
&=& \frac{1-\alpha^2}{2},
\end{eqnarray}
which is the expected result for HOM interference when photons are mismatched and there is no filtering. Specifically, when the photons are indistinguishable, \mbox{$\alpha=1$}, \mbox{$P(M)=0$} and we never observe coincidence events. Whereas for distinguishable photons, \mbox{$\alpha=0$}, \mbox{$P(M)=1/2$} and the photons behave as classical particles.

%
%

\subsection{Spectrally mixed photons}

Thus far, we have considered boson-sampling where each input photon has distinct spectral structure, but all photons are assumed to be spectrally pure. In many real-world experiments, photons are often spectrally mixed. For example, if photons are prepared via heralded spontaneous parametric down-conversion, then spectral correlations between the signal and idler photons may yield mixing in the spectral degree of freedom, depending on the spectral characteristics of the heralding photodetector.

Let each input photon be an arbitrary mixture of $q$ different spectral states,
\begin{equation}
\hat\rho_\mathrm{in} = \bigotimes_{i=1}^n \hat\rho_i,
\end{equation}
where,
\begin{equation} \label{eq:mixed_rho}
\hat\rho_i = \sum_{j=1}^q p_{i,j} \hat{A}_{\psi_{i,j}}^\dag \ket{0}\bra{0} \hat{A}_{\psi_{i,j}},
\end{equation}
where
\mbox{$\psi_{i,j}$} is the spectral distribution function of the $j$th term in the mixture of the $i$th photon.

The probability that for the whole $n$-photon state, the $i$th photon is in the $j_i$th spectral state is given by,
\begin{equation}
\prod_{i=1}^n p_{i,j_i},
\end{equation}
and the respective sampling probability follows from Eq. \ref{eq:general_gamma} as
\begin{equation}P_{\vec\psi}(\vec{S}) \prod_{i=1}^n p_{i,j_i},
\end{equation}
where \mbox{$\vec\psi = \{\psi_{j_1},\dots,\psi_{j_n}\}$}. Now the total probability of measuring a given output configuration is given by summing this expression over all combinations of components in the input mixture,
\begin{equation}
P(\vec{S}) = \sum_{j_1,\dots,j_n=1}^q P_{\vec\psi}(\vec{S}) \prod_{i=1}^n p_{i,j_i},
\end{equation}
which requires summing over $q^n$ amplitudes, which clearly grows exponentially in $n$, except in the trivial case where all photons are spectrally pure and thus contain only a single term in Eq. \ref{eq:mixed_rho}.

%
%

\section{Conclusion}

We have generalized the boson-sampling model to photons of arbitrary spectral structure. We found that while in ordinary boson-sampling with indistinguishable photons each amplitude is related to a matrix permanent, in general the amplitudes are linear combinations of products of functions of permanents. Our result reduces to the usual boson-sampling permanent result when photons are indistinguishable, and also verifies that with distinguishable photons boson-sampling is computationally easy, an expected outcome. Specifically, for indistinguishable photons the measurement probabilities are given by $|\mathrm{Per}(U')|^2$, where $U'$ is a submatrix of $U$, whereas for distinguishable photons they are given by $\mathrm{Per}(|U'|^2)$. The former is computationally hard, whereas the latter is computationally easy to approximate. In the intermediate regime, our results demonstrate the relationship between photonic states and functions of permanents, which sheds light on the question of their computational complexity. Our results apply for general detector models, whereby the detectors are either spectrally-resolving or blind to the spectral structure of photons. We demonstrated that in both detector regimes, our model reproduces expected Hong-Ou-Mandel two-photon interference effects.

%
%

\begin{acknowledgments}
Whilst preparing this manuscript Shchesnovich \cite{bib:Shchesnovich14} presented a similar treatment of the boson-sampling problem with photons of different spectral structures. We refer the reader to Ref. \cite{bib:Shchesnovich14} for details of their approach.

We thank Bill Munro, Timothy Ralph and Barry Sanders for helpful discussions. This research was conducted by the Australian Research Council Centre of Excellence for Engineered Quantum Systems (Project number CE110001013).
\end{acknowledgments}

%
%

\bibliography{bibliography}

\end{document}